\documentclass{ws-procs975x65}

\begin{document}

\title{BREAKDOWN OF THE EQUIVALENCE BETWEEN  \\ PASSIVE
GRAVITATIONAL MASS AND ENERGY \\ FOR A QUANTUM BODY\\
}

\author{ANDREI G. LEBED}

\address{Department of Physics, University of Arizona,\\
Tucson, Arizona 85721, USA and \\
L.D.Landau Institute for Theoretical Physics, \\
Moscow 1117334, Russia \\
E-mail: lebed@physics.arizona.edu\\
www.arizona.edu}

\begin{abstract}
It is shown that passive gravitational mass operator of a hydrogen
atom in the post-Newtonian approximation of the general relativity
does not commute with its energy operator, taken in the absence of
gravitational field. Nevertheless, the equivalence between the
expectation values of passive gravitational mass and energy is shown 
to survive at a macroscopic level for stationary quantum states.
Breakdown of the equivalence between passive gravitational mass and
energy at a microscopic level for stationary quantum states can be
experimentally detected by studying unusual electromagnetic
radiation, emitted by the atoms, supported and moved in the Earth
gravitational field with constant velocity using spacecraft or
satellite.
\end{abstract}


\bodymatter

\section{Introduction}
It is known that gravitational mass of a composite classical body in
the general relativity is not a trivial notion. For example, for two
electrostatically bound objects with bare masses $m_1$ and $m_2$,
only averaged over time gravitational mass, $<m^g>_t$, satisfies the
Einstein equation \cite{Nord,Carlip}:
\begin{equation}
<m^g>_t = m_1 + m_2 + <K + U>_t +<2K + U>_t = m_1 + m_2 +E/c^2 ,
\end{equation}
where $K$ is kinetic energy, $U$ is potential energy, $E$ is the
total energy. It is important that the virial term in Eq. (1) is
zero due to the virial theorem, $<2K + U>_t=0$.

\section{Gravitational mass of a quantum body}
The main goal of our paper in to consider a quantum problem about
passive gravitational mass of a hydrogen atom in the Earth
gravitational field. We define the gravitational mass as a quantity
proportional to a weight of the atom in a weak centrosymmetric
gravitational field \cite{Misner-1},
\begin{equation}
d s^2 = -\biggl(1 + 2 \frac{\phi}{c^2} \biggl)(cdt)^2 + \biggl(1 - 2
\frac{\phi}{c^2} \biggl) (dx^2 +dy^2+dz^2 ), \ \phi = - \frac{GM}{R}
,
\end{equation}
where $|\phi|/c^2 \ll 1$, $G$ is the gravitational constant, $c$ is
the velocity of light, $M$ is the Earth mass, $R$ is a distance from
a center of the Earth and a center of mass of a hydrogen atom (i.e.,
proton).

For interval in Eq. (2), the effective Schrodinger equation for a
hydrogen atom is derived in Ref.~\citen{Fisch} to calculate the
so-called "gravitational Start effect". Below, we consider
completely different phenomena \cite{Lebed} and have to use the
Hamiltonian from Eq. (3.24) of Ref.~\citen{Fisch} without tidal
terms. As a result of disregarding all tidal terms, we obtain the
following Schrodinger equation:
\begin{equation}
\hat H = m_e c^2 + \frac{\hat {\bf p}^2}{2m_e}-\frac{e^2}{r} + \hat
m^g_e \phi \ ,
\end{equation}
where we introduce passive gravitational mass operator of an
electron as
\begin{equation}
\hat m^g_e  = m_e + \biggl(\frac{\hat {\bf p}^2}{2m_e}
-\frac{e^2}{r}\biggl)\frac{1}{c^2} + \biggl(2 \frac{\hat {\bf
p}^2}{2m_e}-\frac{e^2}{r} \biggl) \frac{1}{c^2} \ ,
\end{equation}
where $m_e$ is a bare electron mass, $\hat{\bf p}$ is electron
momentum operator, $r$ is a distance between electron and proton.
Suppose that we have macroscopic ensemble of hydrogen atoms with
each of them being in a ground state with energy $E_1$. Then, as
follows from Eq. (4), the expectation value of the gravitational
mass per one electron is:
\begin{equation}
<\hat m^g_e> = m_e + \frac{ E_1}{c^2}  + \biggl< 2 \frac{\hat
{\bf p}^2}{2m_e}-\frac{e^2}{r} \biggl> \frac{1}{c^2} = m_e +
\frac{E_1}{c^2}  ,
\end{equation}
where the term in brackets is zero due to the quantum virial theorem
\cite{Park}. Thus, we formulate our first result: the equivalence
between passive gravitational mass and energy, taken in the absence
of gravitational field, survives at a macroscopic level for
stationary quantum states \cite{Lebed}.

Here, we describe a thought experiment, which shows that Eq. (4)
breaks the equivalence between passive gravitational mass and energy
at a microscopic level, which is our second result \cite{Lebed}. We
consider the case, where gravitational field is adiabatically
switched on, which corresponds to the following time-dependent
perturbation:
\begin{equation}
\hat U ({\bf r},t) = \phi(R) \exp(\lambda t) \biggl[
\biggl(\frac{\hat {\bf p}^2}{2m_e}-\frac{e^2}{r}\biggl)/c^2 +
\biggl(2 \frac{\hat {\bf p}^2}{2m_e}-\frac{e^2}{r} \biggl)/ c^2
\biggl] , \ \lambda \rightarrow 0.
\end{equation}
[Note that our choice of the perturbation in Eq. (6) allows to
disregard all velocity dependent terms.] Suppose that, at $t
\rightarrow - \infty$ (i.e., in the absence of the field), a
hydrogen atom is in its ground state,
\begin{equation}
\Psi_1(r,t) = \Psi_1(r) \exp(-iE_1t/\hbar) \ ,
\end{equation}
Then, at $t \rightarrow 0$ (i.e., in the presence of the field), the electron
wave function can be written as
\begin{equation}
\Psi(r,t) = \sum^{\infty}_{n = 1} a_n(t) \Psi_n(r) \exp(-i E_n t/\hbar)
\ ,
\end{equation}
where $\Psi_n(r)$ are normalized s-type electron wave functions with
energies $E_n$. The standard calculations show that the probability
that, at $t=0$, an electron occupies $n$-th $(n \neq 1)$ energy level
is
\begin{equation}
P_n = |a_n(0)|^2=[\phi(R)/c^2]^2 \ [V_{n,1}/(E_n-E_1)]^2  \simeq 0.49
\times 10^{-18} [V_{n,1}/(E_n-E_1)]^2,
\end{equation}
where $V_{n,1}$ is a matrix element of the virial operator, $\hat
V({\bf r}) = 2 \frac{\hat {\bf p}^2}{2m_e}-\frac{e^2}{r}$,
\begin{equation}
V_{n,1}= \int \Psi^*_n(r) \hat V({\bf r}) \Psi_1(r) d^3 {\bf r} , \
\hbar \omega_{n,1}= (E_n-E_1)/\hbar , \ n \neq 1,
\end{equation}
[Here, we use $M \simeq 6 \times 10^{24}$kg, $R \simeq 6.4 \times
10^6$m.] In fact, this means that a measurement of the gravitational
mass (4) gives the following quantized values:
\begin{equation}
m^g_e (n) = m_e + E_n/c^2 \ ,
\end{equation}
instead of the expected Einstein equation, $m^g_e=m_e+E_1/c^2$. It
is important that the excited energy levels spontaneously decay and,
thus, the quantization law of Eq. (11) can be detected by measuring
electromagnetic radiation, emitted by macroscopic ensemble of
hydrogen atoms.

\section{Suggested experiment}

Let us suggest a realistic experiment to detect the above discussed
inequivalence, which is our third result \cite{Lebed}. Suppose that
a hydrogen atom is in its ground state at $t=0$ and located at
position $R$. Suppose that it is moved with constant velocity
using spacecraft or satellite. The corresponding time dependent
perturbation is
\begin{equation}
\hat U ({\bf r},t) =\frac{\phi(R+vt)-\phi(R)}{c^2} \biggl(3
\frac{\hat {\bf p}^2}{2m_e}-2\frac{e^2}{r} \biggl).
\end{equation}
It is possible to show that, for the case $|\phi(R+vt)| \ll
|\phi(R)|$, the probability that an electron is excited on $n$-th
energy level coincides with Eq. (9). Note that, although the
probabilities in Eq. (9) are small, the number of photons, emitted
by macroscopic ensemble of the atoms, can be large. For 1000 mole of
hydrogen atoms, the number of photons, emitted with frequency
$\omega_{2,1} = (E_2 - E_1)/\hbar$ is
\begin{equation}
N(2 \rightarrow 1) = 0.9 \times 10^8 .
\end{equation}
To the best of our knowledge, if such experiment is done, it will be
the first direct test of some nontrivial combination of the general
relativity and quantum mechanics.

\section*{Acknowledgments} We are thankful to N.N. Bagmet (Lebed), V.A. Belinski,
and Li Zhi Fang for useful discussions. This work is supported by
the NSF under Grant DMR-1104512.


\bibliographystyle{ws-procs975x65}
\bibliography{ws-pro-sample}

\end{document}